\documentclass[doublecol]{epl2} 

\usepackage{amsmath}
\usepackage{amssymb}
\usepackage{amsfonts}

\newcommand{\veci}[1]{\bm{#1}_i}
\newcommand{\vecidot}[1]{\dot{\bm{#1}}_i}

\title{Spontaneous velocity alignment of Brownian particles with feedback-induced propulsion}
\shorttitle{Spontaneous velocity alignment of Brownian particles with feedback-induced propulsion}
\author{R. A. Kopp\inst{1} \and S. H. L. Klapp\inst{1}}
\institute{                    
	\inst{1} Technische Universität Berlin, Institut für Theoretische Physik - Hardenbergstr. 36, 10623 Berlin 
}
\abstract{
Based on Brownian dynamics simulations we study the collective behavior of a two-dimensional system of repulsively interacting colloidal particles,
where each particle is propelled by a repulsive feedback force with time delay $\tau$. Although the pair interactions are purely isotropic we observe
a spontaneous, large-scale alignment of the velocity vectors. This phenomenon persists for long times and occurs in the absence of steady-state clustering. 
We explain our observations by a combination of the effect of steric interactions yielding local velocity ordering, and the effect of time delay, that generates cluster dissolution, velocity persistence 
and velocity alignment over large distances.
Overall, the behavior reveals intriguing similarities, but also differences, to that observed in models of active matter, such as active Brownian particles and the Vicsek model.
}
\begin{document}
\maketitle
\section{Introduction}
In recent years the behavior of soft-matter and biological systems involving non-Markovian (memory) effects and feedback has become a focus of growing interest.
Indeed, non-Markovian effects
play an important role in many autonomous biological and neural systems where there is some time lag (or {\em time delay}) between the reception of information and the actual response, examples being sensorial delay in robotics \cite{mijalkov_engineering_2016}, communication delay in swarms \cite{hindes_hybrid_2016}, or delayed motion of cells due to a viscoelastic substrate \cite{clark_self-generated_2022}. Yet another example is inertia-induced delay which has recently received much attention in the context of active systems \cite{lowen_inertial_2020}.
Besides natural systems, time delay is a characteristic (and often unavoidable) feature of many artificial feedback {\em control} loops affecting, e.g., the position or orientation of a colloid, which can be realized, e.g., via photon nudging or adaptive light fields \cite{qian_harnessing_2013,mijalkov_engineering_2016,franzl_active_2020, fernandez-rodriguez_feedback-controlled_2020,loffler_behavior-dependent_2021}. Using such protocols for colloids is a topic receiving strong interest in view of potential applications (e.g., optimization of transport \cite{gernert_enhancement_2015}) but also from a fundamental perspective in terms of pattern formation \cite{tarama_traveling_2019} and thermodynamics
\cite{toyabe_experimental_2010,blickle_realization_2012,koski_experimental_2014,jun_high-precision_2014,loos_irreversibility_2020,cao_thermodynamics_2009}, particularly when combined with time delay \cite{debiossac_thermodynamics_2020, loos_irreversibility_2020, loos_heat_2019,munakata_entropy_2014}.
Moreover, recent studies of time-delayed feedback control of {\em active} colloids \cite{franzl_active_2020} have revealed new effects such as delay-induced clustering and swarming \cite{mijalkov_engineering_2016,holubec_finite-size_2021-1,sprenger_active_2022,wang_spontaneous_2023}, enriching the already complex collective behavior of active matter. However, so far, there is no comprehensive understanding of the role of time delay and its interplay with activity on structure formation and emergent dynamics of interacting, Brownian systems.

As a step in this direction we here discuss a colloidal system where activity is {\em induced} by time delay. Specifically, our model system is composed of repulsively interacting (isotropic) particles subject to nonlinear, time-delayed repulsive feedback.
In a previous study \cite{kopp_persistent_2023-1} on the single-particle dynamics we have shown that the time delay can actually generate {\em persistent} motion: at appropriate feedback parameters, the particle develops a constant effective velocity $v_{\mathrm{eff}}$ whose direction randomizes after some persistence time $\tau_\mathrm{eff}$. Overall, the single-particle dynamics reveals crucial similarities 
to that of active Brownian particles (ABP) \cite{bechinger_active_2016}, suggesting that time-delayed feedback could indeed be an alternative mechanism to induce self-propulsion.
Therefore, and given the wealth of knowledge on the collective behavior of ABPs (and related models of active particles \cite{caprini_spatial_2021-1}), including motility-induced phase separation (MIPS) \cite{cates_motility-induced_2015} and local velocity correlations \cite{caprini_spontaneous_2020}, it is an intriguing question to which end the delay-driven particles mimic this behavior (or not).

To this end we here present results from extensive Brownian dynamics (BD) simulations in two dimensions (2D).
 Unexpectedly, we find that the persistent motion of the individual, delay-driven particles can synchronize, yielding velocity ordered states. This is accompanied by a {\em transient} formation of clusters which, at high densities, develops into an essentially homogeneous particle distribution. We rationalize this behavior as a combination of local alignment generated by steric interactions (as has been seen already in systems of ABPs \cite{caprini_spontaneous_2020}), and specific delay effects enabling the transport of alignment information and, thus, ordering, on larger scales.
\section{Model and methods}
We simulate a 2D system of $N=3200$ interacting particles in a quadratic box with periodic boundary conditions. Each particle $i=1,\ldots,N$ with position $\veci{r}(t)$ moves according the overdamped Langevin equation
\begin{equation}
	\gamma \vecidot{r} = \bm{F}_i^\mathrm{FB}\left(\veci{r}(t),\veci{r}(t-\tau)\right)+ \bm{F}_i^\mathrm{WCA}+\bm{\xi}_i(t),
	\label{eq:BD}
\end{equation}
where $\gamma$ is the friction constant and $\bm{\xi}_i(t)$ is a Gaussian white noise with correlation function $\langle \xi_{i,\alpha} (t) \xi_{j,\beta}(t^\prime) \rangle = 2 \gamma^2 D_0\delta_{ij}\delta_{\alpha \beta} \delta(t-t^\prime)$ (with $D_0$ being the translational diffusion constant given by the Einstein relation, i.e., $D_0=k_BT/\gamma$, 
where $k_B$ and $T$ are Boltzmann's constant and the temperature, respectively.
The first term on the right side of eq.~(\ref{eq:BD}) represents a repulsive, time-delayed feedback force acting on each particle $i$ individually. Specifically,
$\bm{F}_i^\mathrm{FB}(\veci{R}) = -\bm{\nabla}_i U^{FB}(|\veci{R}|)$ where $\veci{R}=\veci{r}(t)-\veci{r}(t-\tau)$ is the displacement within one delay time $\tau$ and
$U^{\mathrm{FB}}(|\veci{R}|)= A\exp(-\veci{R}^2/2b^2)$. The feedback strength $A$ and feedback range $b$ are set to positive constants. Physically, 
the feedback potential can then be interpreted as a source of a ``nudge'' following the particle with some delay \cite{kopp_persistent_2023-1}.  Such a potential could, in principle, be created by optical forces  \cite{leyman_tuning_2018,lavergne_group_2019,franzl_active_2020,loffler_behavior-dependent_2021}; similar time-delayed potentials occur, e.g., when cells move in a viscoelastic medium \cite{clark_self-generated_2022}.
As shown in \cite{kopp_persistent_2023-1} for the single-particle case, feedback parameters fulfilling the condition $A\tau/ \gamma b^2>1$
lead to a constant, long-time velocity vector $\bm{v}_{i,\infty}$ of the particle in the deterministic limit. The magnitude of this velocity can be obtained analytically, it is given by 
$v_\infty= (\sqrt{2}b/\tau)\sqrt{-\ln\left(\gamma b^2/(A\tau)\right)}$ \cite{kopp_persistent_2023-1}.
In the presence of noise, the direction of the single-particle velocity randomizes on the time scale of a ``persistence time'' $\tau_{\mathrm{eff}}$
while the finite magnitude $v_{\mathrm{eff}}$ remains, on average, constant in the long-time limit (with a value close to $v_\infty$). The resulting mean-squared displacement (MSD) resembles \cite{kopp_persistent_2023-1} 
that of an ABP (see supplementary information (SI) \cite{kopp_supplemental_2023} for an example).
The particles interact via the (instantaneous) repulsive forces $\bm{F}^{\mathrm{WCA}}_{i} = -\sum_{j\neq i}\nabla_{\bm{r}_i}U_{\mathrm{WCA}}(r_{ij})$ where $U_{\mathrm{WCA}}$ is the standard Weeks-Chandler-Andersen (WCA) potential \cite{weeks_role_1971}. Throughout the paper, the interaction strength is set to $\varepsilon/k_B T = 100$
and the cutoff radius $r_c = 2^{1/6} \sigma$. We take this length as the effective diameter of our particles of radius $R$, that is, $r_c=2R$, which is also used as a length scale.
The same length is used for the range of the feedback potential, i.e., $b=2R$.
Further, times are expressed via the Brownian time $\tau_B = \left(2R\right)^2/D_0$.
The system's behavior is then determined by three dimensionless parameters:  area packing fraction $\phi=N\pi R^2/L^2$, feedback strengths $A/k_B T$, and delay time $\tau/\tau_B$. Clearly, for $\phi\to 0$ the dynamics reduces to that of a single particle \cite{kopp_persistent_2023-1}. On the other hand, the limits $A/k_B T\to 0$, $\tau\to 0$ (or $b\to 0$) all lead to a passive system of particles interacting only via WCA forces.
Here, we mostly choose (if not stated otherwise) $\tau=0.15 \tau_B$, which is in the upper range of typical values of $\tau$ in experiments \cite{franzl_active_2020, lopez_realization_2008}.
The equations of motion are integrated using the Euler-Maruyama integration scheme \cite{maruyama_continuous_1955} with a time step $\delta t = 10^{-5}\tau_B$. To account for the periodic boundary conditions, we use the minimal image convention. Thereby, we choose the box length $L$ such that a particle never travels further than $L/2$ within one $\tau_B$ (and, thus, several $\tau$). Due to the time-delayed character of $\bm{F}_i^\mathrm{FB}$, one has to define for each particle a history function before switching on the feedback at $t=0\tau_B$. To this end, we utilize the stochastic trajectories of ``passive'' Brownian particles interacting via the WCA potential alone.

\section{Dynamical behavior}
\label{sec:dynamics_feedback}
\begin{figure}
\includegraphics[width=0.49\textwidth]{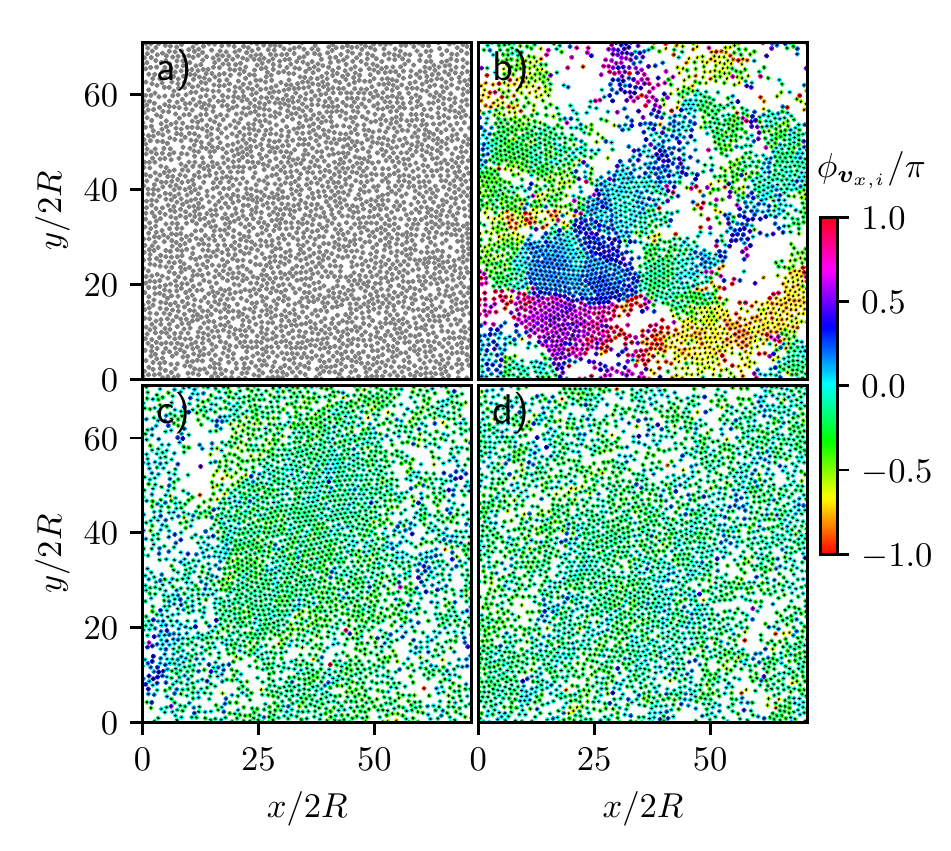}
\caption{Simulation snapshots at different points in time. Part a) pertains to a time before the onset of feedback ($t=0$), where the system is passive. Parts b)-d) pertain to time steps after onset of feedback  at b) $t = 2.5\tau_B$, c) $t = 12.5\tau_B$, d) $t = 77.5\tau_B$. In parts b)-d), particles are colored according to the direction of their velocities relative to the $x$-axis.
Parameters: $\phi = 0.5$, $A = 140k_B T$, $b = 2R$, $\tau = 0.15 \tau_B$.}
\label{fig:snapshot}
\end{figure}

We start the discussion directly with the main observation of our numerical simulations, that is, the emergence of a state with aligned velocity vectors of the particles. To this end we consider a relatively dense system ($\phi = 0.5$)
with strong repulsive feedback ($A/k_BT=140$). An illustration of the behavior of the dense, interacting system is given in fig.~\ref{fig:snapshot}, where we show simulation snapshots at different points in time. A corresponding movie (Supplementary Movie 1) is provided in the supplementary material \cite{kopp_supplemental_2023}. 
Before the onset of the feedback forces, when the particles are ``passive'' and just interact via steric forces, the system is essentially homogeneous (see fig.~\ref{fig:snapshot}a)). Switching on the feedback,
each particle starts to perform persistent random motion until it hits other particles. As seen from fig.~\ref{fig:snapshot}b) this leads, first, to the formation of clusters, a behavior not seen in the corresponding passive system. It is well known, however, that clustering does occur in dense non-equilibrium systems of active particles, such as ABPs \cite{zottl_emergent_2016-1}. Despite this apparent similarity, a major difference occurs when we consider the further progress in time:
In the feedback-driven system, the clusters do {\em not} persist, rather they dissolve in the course of time (figs.~\ref{fig:snapshot}c) and d)). As a consequence, the spatial structure at long times appears quite homogeneous (fig.~\ref{fig:snapshot}d)) at $\phi=0.5$.  
This behavior is in marked contrast to that of a system of active particles, particularly ABPs, as we will later show (see fig.~\ref{fig:ABP}).

\begin{figure}
\includegraphics[width=0.24\textwidth]{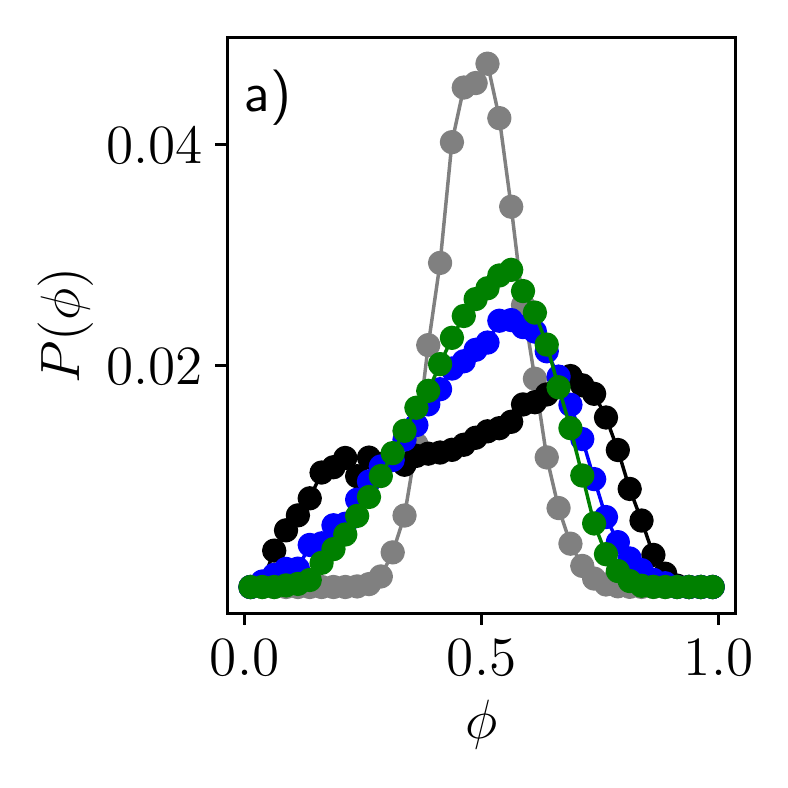}
\includegraphics[width=0.24\textwidth]{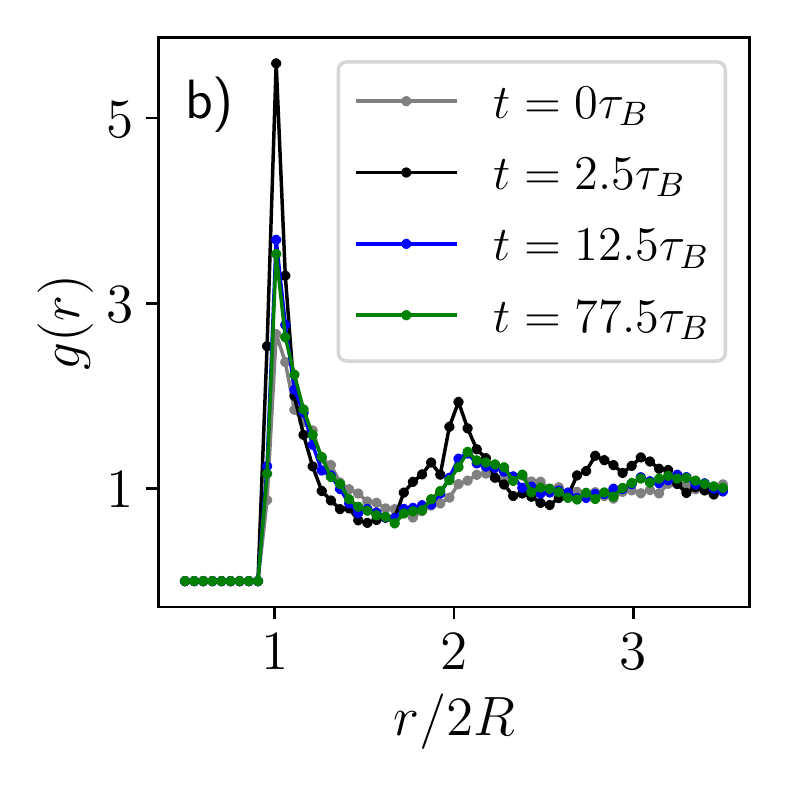}
\includegraphics[width=0.24\textwidth]{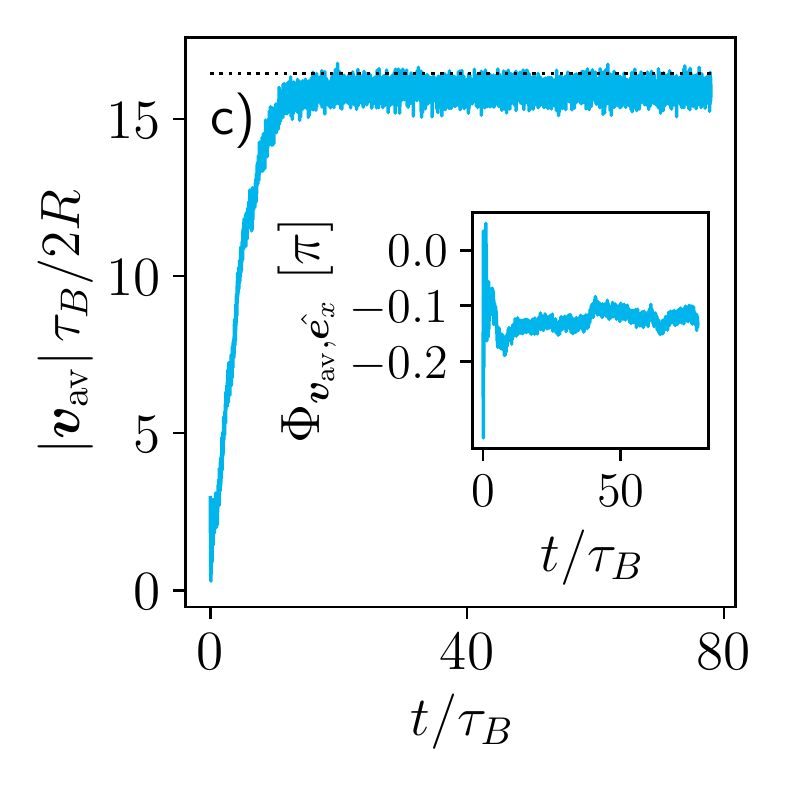}
\includegraphics[width=0.24\textwidth]{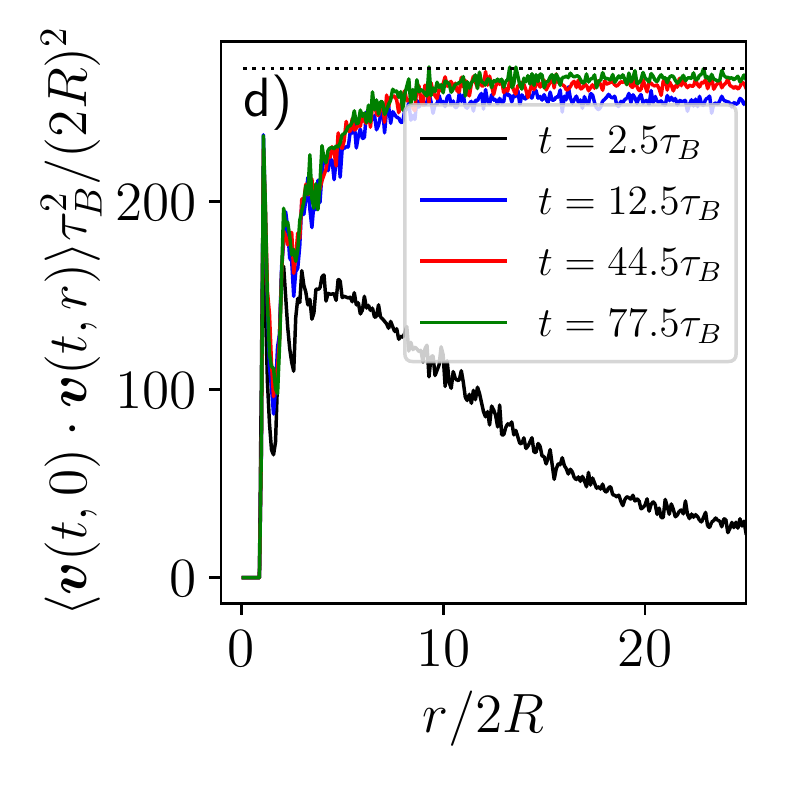}
\caption{a) Distribution of local area fractions at several points in time (see legend in panel b)). b) Radial distribution function at several points in time.
c) Magnitude of the net velocity, $v_{\mathrm{av}}$, as function of time for a single realization of noise. The dashed horizontal line
indicates the long-time velocity of a single particle in the deterministic case. Inset:
Angle of $\bm{v}_{\mathrm{av}}$ relative to $x$-axis (in units of $\pi$). d) Averaged spatial velocity correlation functions at different points in time (parameters as in fig.~\ref{fig:snapshot}).}
\label{fig:structure}
\end{figure}

The appearance of {\em transient} clustering in the present system is also visible from the distributions of local area fractions, $P(\phi)$ \cite{kopp_supplemental_2023}, at several points in time, see fig.~\ref{fig:structure}a).
Starting from a pronounced maximum of $P(\phi)$ at the average density $\phi = 0.5$ in the ``passive'' case ($t=0$), the maximum first becomes substantially broader, indicating the presence of different local area fractions, and, thus, clustering. At first sight, the curve at $t = 2.5\tau_B$ points to the emergence of a double-peak structure in $P(\phi)$, as it is familiar from ABP systems close to MIPS. However, as time proceeds ($t\gtrsim 12.5\tau_B$), the distribution narrows while the maximum grows, indicating increasing homogenization of the system.
These trends are also reflected by the 2D pair correlation function $g(r)$, whose structure becomes increasingly softer when time proceeds (see fig.~\ref{fig:structure}b)).

We understand the clustering behavior as follows. After the onset of feedback, each particle develops persistent motion in a random direction and thereby behaves, at first, similar to an active particle. In particular, if the density is sufficiently high, the particles collide with other particles and form, at first, moving clusters.
However, the hindrance of motion due to collisions
leads to smaller displacements $\bm{r}_i(t)-\bm{r}_i(t-\tau)$. Thereby, the collisions effectively decrease the feedback force and thus, lead to a smaller velocity of the involved particles in the next time interval of length $\tau$.
When the velocities become too small (or the particles even stop moving at all), the clusters dissolve. Only at somewhat later times (beyond the next delay interval) each particle has ``forgotten'' the hindering effect, such that, in principle, new clusters can form.

We now turn to the dynamics of the particle velocity vectors, $\veci{v}(t)=\vecidot{r}(t)$. In figs.~\ref{fig:snapshot}b)-d), the instantaneous directions of $\veci{v}$ relative to the $x$-axis are indicated by the color bar.
At times pertaining to clustering, the velocities within the clusters start to align, as seen from the equal-colored regions in fig.~\ref{fig:snapshot}b). Intriguingly, as time proceeds, the velocities align more and more, while the clusters themselves dissolve (see Supplementary Movie 1 \cite{kopp_supplemental_2023}). To quantify the degree of the emergent ``velocity alignment'', we calculate the magnitude of the system-averaged velocity,
$v_{\mathrm{av}}=|\bm{v}_\mathrm{av}|=|N^{-1}\sum_i\veci{v}(t)|$. The time dependence of this quantity for a single realization of noise is shown in 
fig.~\ref{fig:structure}c). Despite strong fluctuations (which are expected in an overdamped system) one observes 
that $v_{\mathrm{av}}$ approaches a constant value at long times (consistent with that found from the (noise-averaged) many-particle MSD (see SI \cite{kopp_supplemental_2023}).
Further, along with the magnitude, also the polar angle of the velocity {\em vector}, $\bm{v}_{\mathrm{av}}=N^{-1}\sum_i\veci{v}(t)$, settles in time and then fluctuates only slowly around a constant value, as shown in the inset of 
fig.~\ref{fig:structure}c).
Both features clearly point to the emergence of a steady state characterized by large-scale velocity alignment.
Interestingly, the saturation value of $v_{\mathrm{av}}$ is close to, yet 
somewhat smaller than the velocity of a {\em single}, deterministic feedback-driven particle \cite{kopp_persistent_2023-1} at the same parameters ($v_{\infty}\tau_B/2R\approx 16.45$). This reflects a certain degree of orientational friction induced by neighboring particles.

As an additional indicator of large-scale velocity ordering we have plotted in 
fig.~\ref{fig:structure}d) the (system-averaged) velocity correlations as function of the distance between pairs of particles at several points in time. Similar correlation functions have been discussed for {\em active} particle systems \cite{caprini_spatial_2021-1}. Figure~\ref{fig:structure}d)  clearly shows how the range of the velocity correlations grows as time proceeds. The smallest time ($t=2.5\tau_B$) considered in 
fig.~\ref{fig:structure}d) 
pertains to transient clustering.
Already at this short time, the correlations extend over more than ten particle diameters before they decay to zero. In contrast, at later times, the correlation function approaches a constant value, which is consistent with the square of the long-time limit of $v_{\mathrm{av}}$.
\section{Comparison to active Brownian particles}
\label{sec:dynamics_active}
The collective behavior of the feedback-driven system displays similarities, but also differences to what one observes in active systems. To further explore this point,
we consider a system of ABPs with purely repulsive (WCA) interactions and activity parameters comparable to that of a feedback-driven system at $A/k_B T=140$ (see SI \cite{kopp_supplemental_2023} for details).

\begin{figure}
\includegraphics[width=0.24\textwidth]{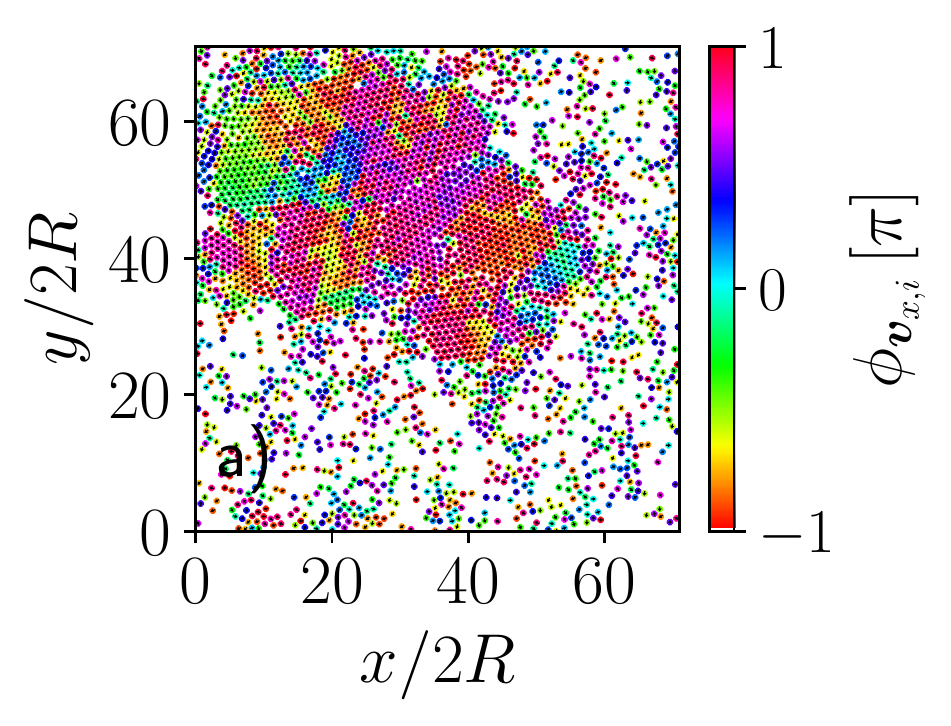}
\includegraphics[width=0.24\textwidth]{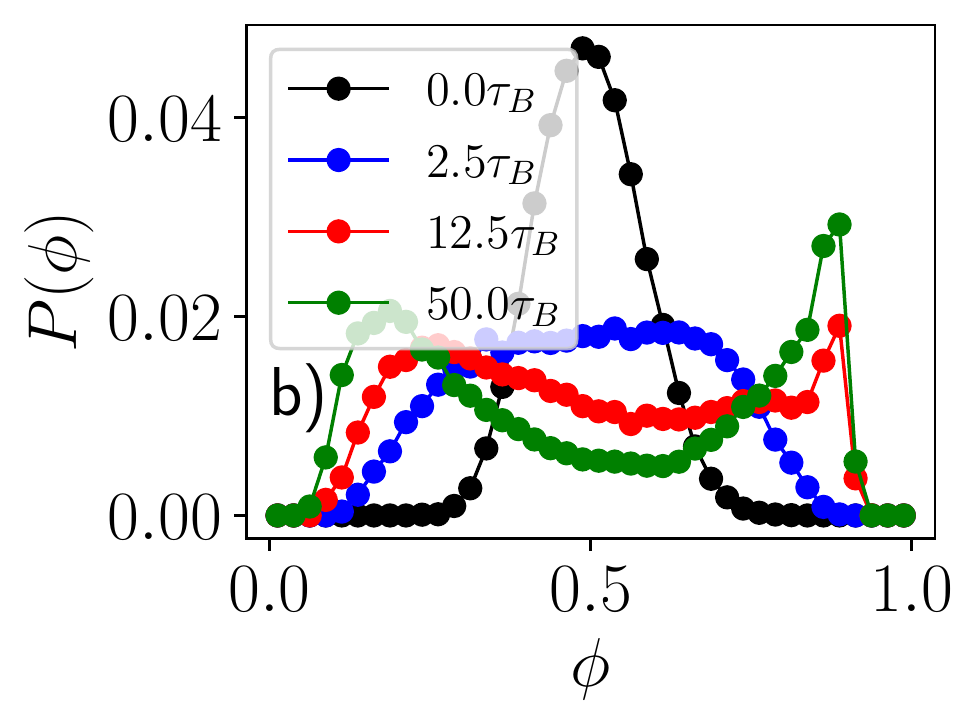}	
	\caption{a) Simulation snapshot at $t=50\tau_B$ (with particles colored according to the direction of their velocities relative to the $x$-axis)
and b) distribution of local area fraction in a system of active Brownian particles at $\phi=0.5$ with propulsion speed $v_0 \tau_B/2R=16.96$ and persistence time $\tau_r=1/D_r=1.577 \tau_B$, corresponding to the parameters of the feedback-driven system in fig.~\ref{fig:snapshot}.}
	\label{fig:ABP}
\end{figure}

The (long-time) behavior of the resulting ABP system at $\phi=0.5$ as obtained from BD simulations (with WCA parameter chosen as before) is illustrated in fig.~\ref{fig:ABP}.
As a major difference to the feedback-driven system (cf. fig.~\ref{fig:snapshot}), we see from fig.~\ref{fig:ABP}a) that the ABP system displays {\em stable}, large clusters, accompanied by a bimodal structure of $P(\phi)$ (fig.~\ref{fig:ABP}b)). We take this as a hint that the here considered ABP system exhibits MIPS, a well-known phenomenon in this type of active matter \cite{cates_motility-induced_2015}. The occurrence of MIPS is one of the major differences to the present, feedback-driven system.
Moreover, as seen from the color code in fig.~\ref{fig:ABP}a),
the ABPs align their velocities only {\em locally} (consistent with \cite{caprini_spontaneous_2020}). This is is stark contrast to the large-scale ordering seen in the present system.

\section{State diagram}
\label{sec:state}
\begin{figure}
	\onefigure[width=0.49\textwidth]{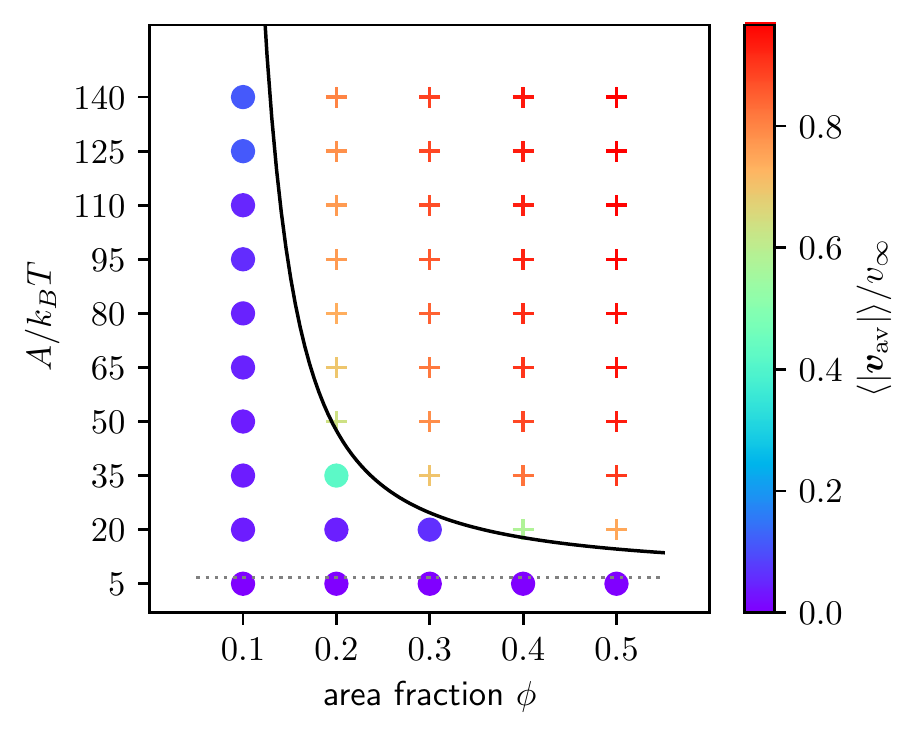}
	\caption{State diagram evaluated in the long-time limit ($t=75\tau_B$). Circles: disordered states, crosses: velocity-aligned states, with long-time velocity magnitude (normalized by the single-particle value in the deterministic limit, $v_\infty$) given by the color bar. The black solid line represents our mean-field estimate for the onset of ordering, eq.~(\ref{eq:onset}).
	The horizontal dashed line indicates the lower threshold of $A/k_B T$ for the onset of motion in the (deterministic) single-particle case. For smaller feedback strength, $v_\infty$ does not exist, and we have put the BD data points to zero.}.
	\label{fig:state_diag}
\end{figure}

So far we have focussed on a particular value of $\phi$ and $A/k_B T$. An overview of the occurrence of velocity ordering in the plane spanned by these parameters is given in fig.~\ref{fig:state_diag}.
The color bar on the right side shows the noise-averaged magnitude of the system-averaged velocity, $\langle |\bm{v}_{\mathrm{av}}|\rangle$, relative to the (analytically known) 
velocity $v_{\infty}$ of a single particle in the deterministic case. The latter is non-zero 
for all $A\tau/\gamma b^2>1$ \cite{kopp_persistent_2023-1}, yielding (in the present units and with our choices of $\tau$ and $b$)
$A^\mathrm{min}/k_B T=\tau_B/\tau\approx6.67$.
This value is indicated by the dashed line in fig.~\ref{fig:state_diag}.
We define the interacting system to be velocity-aligned if the resulting normalized velocity $\langle |\bm{v}_{\mathrm{av}}|\rangle/v_\infty$ (whose maximum is $1$) exceeds the value $0.5$ in the long-time limit (for the typical behavior of $v_{\mathrm{av}}$ as function of $A/k_B T$, see SI \cite{kopp_supplemental_2023}). 
The corresponding parameter combinations are indicated by crosses
in fig.~\ref{fig:state_diag}. Closely below the crosses, global alignment is only reached for very long simulation times (see \cite{kopp_supplemental_2023} for details). 
At even smaller $A/k_B T$ (yet above $A^\mathrm{min}/k_B T$) there are, at most, locally aligned patches.

The state diagram reveals that the velocity ordered regime can be reached either by increasing, at fixed $\phi$, the feedback strength $A/k_B T$ from zero (passive limit) to large values, 
or by increasing $\phi$ from zero at fixed $A/k_B T$. Below we derive, based on simple, mean-field-like arguments, an estimate for the feedback strength at the onset of ordering as function of density (or $\phi$) [see eq.~(\ref{eq:onset})].
This function, shown as a solid line in fig.~\ref{fig:state_diag}, describes our numerical data quite well. Consistent with the mean-field prediction, the numerical data further reveal that the larger $\phi$, the closer the onset of ordering moves to $A^\mathrm{min}/k_B T$ (dashed line). Simulations at $\phi=0.6$ (not shown) confirm that this is indeed the case.
The other limit ($\phi\rightarrow 0$) is computationally challenging due to strongly increasing simulation times before reaching the steady state. 
For example, at $\phi=0.1$, ordering occurs only after $\approx 90\tau_B$ at $A/k_BT\approx 180$, that is, if the feedback is increased to values beyond those shown in fig.~\ref{fig:state_diag} \cite{kopp_supplemental_2023}. Still, this indicates that ordering occurs at \emph{any} finite density, if the feedback is sufficiently strong.

Overall, the dependence of velocity ordering on $\phi$ and $A/k_B T$ resembles, when interpreting $A/k_BT$ as inverse noise strength, that of other models of velocity-aligning systems, particularly the paradigmatic Vicsek model \cite{vicsek_novel_1995, chate_collective_2008,romanczuk_active_2012-1,martin-gomez_collective_2018}. In the Vicsek model a characteristic feature of the symmetry-broken state close to the transition points
 is the appearance of \emph{bands}, that is, localized, traveling, high-density, and high-order structures within a less ordered background \cite{chate_collective_2008}.
In our simulations, we have also observed band-like structures, particularly at low densities (see SI \cite{kopp_supplemental_2023} and Supplementary Movie 2 for an example at $\phi=0.2$) and, less pronounced, at higher densities and feedback close above the transition. A more detailed investigation of the character and stability of the bands is, however, beyond the scope of this study.

To complete the overview of the state behavior, we have performed calculations at various other values of the delay time $\tau$ and the feedback range $b$, see SI \cite{kopp_supplemental_2023}.
Starting from a parameter set within the velocity-aligned regime, we found that $\langle |\bm{v}_{\mathrm{av}}|\rangle$  increases if we decrease $\tau$ or increase $b$. Both effects are consistent with the behavior of $v_\infty$ in the single-particle case \cite{kopp_persistent_2023-1}.

\section{Mechanism of local alignment}
\begin{figure}
	\includegraphics[width=0.24\textwidth]{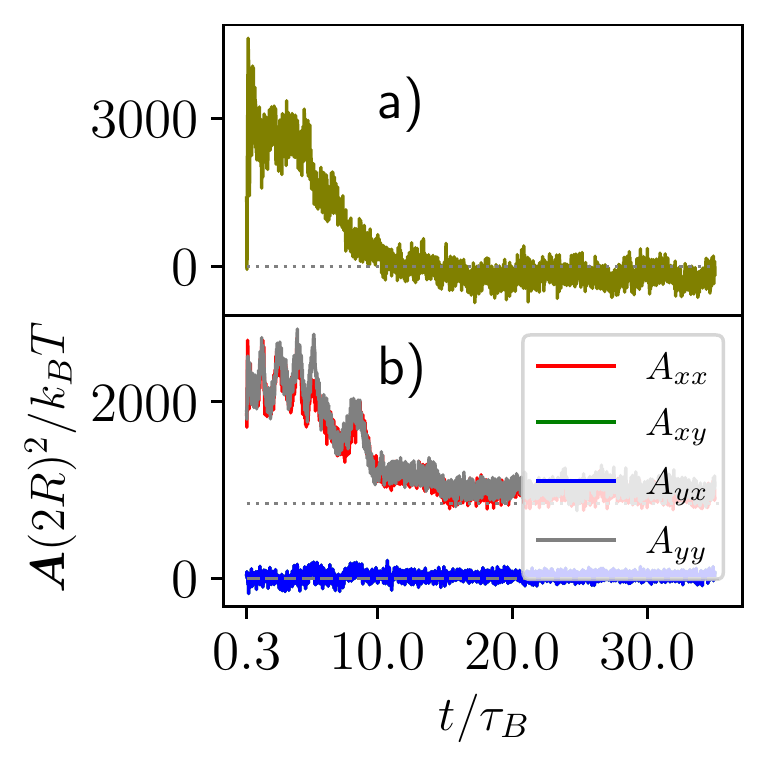}
       \includegraphics[width=0.24\textwidth]{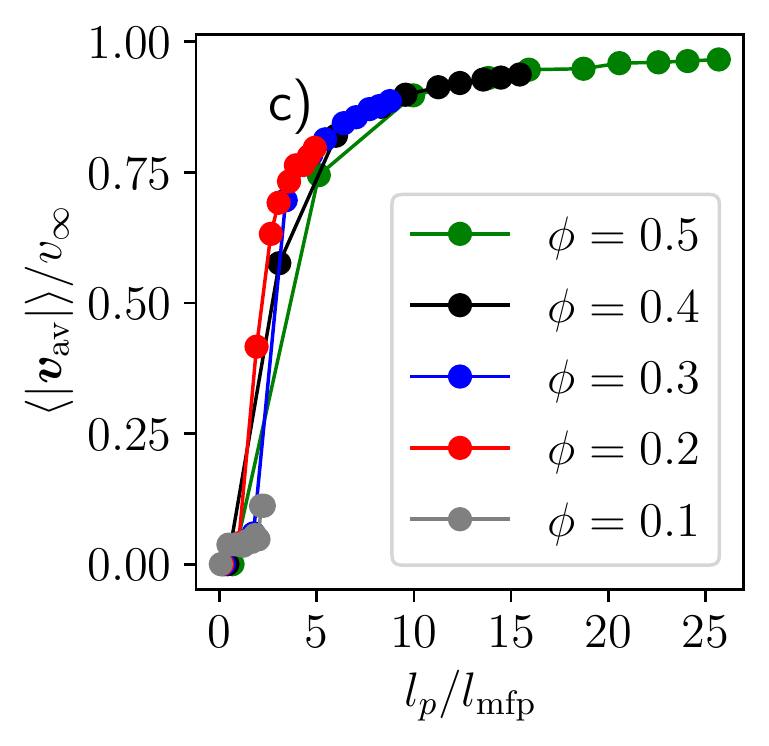}
	\caption{a) System-averaged projection of the first term appearing on the right side of eq.~(\ref{eq:vdot}) onto the individual velocity vector $\bm{v}_i$. 
b) System-averaged elements of the matrices $\bm{A}_i$. The dotted line shows the corresponding mean-field result. All data in a) and b) start at $t=2\tau=0.3\tau_B$ and correspond to a single realization of noise. 
c) Normalized magnitude of the average velocity as function of the ratio of persistence length $l_p$ and mean free path $l_\mathrm{mfp}$.}
	\label{fig:mechanism}
\end{figure}

Even the appearance of \emph{local} velocity ordering between neighbors is not trivial given that there is no direct velocity coupling in eq.~(\ref{eq:BD}).
To understand the effect, we consider in more detail the terms governing the velocity dynamics in eq.~(\ref{eq:BD}).
Specifically, we consider the time derivative of $\bm{v}_i$ in the deterministic limit ($\bm{\xi}(t)=0$), thereby following some steps of a corresponding analysis
done for active Brownian particles\cite{caprini_spontaneous_2020,caprini_spatial_2021-1}. 
From eq.~(\ref{eq:BD}) it then follows \cite{kopp_supplemental_2023} 
\begin{eqnarray}
	\gamma \vecidot{v} &=& -\bm{E}(\bm{R}_i)\left(\bm{v}_i(t) - \bm{v}_i(t-\tau)\right)\nonumber\\
& &- \sum_{j\neq i}\bm{A}(\bm{r}_{ij})\left(\bm{v}_i(t) -\bm{v}_j(t)\right).	
	\label{eq:vdot}
\end{eqnarray}
The first term on the r.h.s. of eq.~(\ref{eq:vdot}) is the feedback (single-particle) contribution, it involves the matrix $\bm{E}(\bm{R}_i)$  with configuration-dependent elements
 $E_{i,\alpha\beta}=-\partial F^{\mathrm{FB}}_{i,\alpha}/\partial R_{i,\beta}$ ($\alpha,\beta=x,y$) \cite{kopp_supplemental_2023}.
 %
 %
This term couples the change of $\veci{v}$ with the difference vector $\bm{v}_i(t) - \bm{v}_i(t-\tau)$; it thus involves the history of the velocity dynamics.
For sufficiently large displacement within one delay time, $\veci{R}=\veci{r}(t)-\veci{r}(t-\tau)$ (i.e., sufficiently strong feedback), $\bm{E}$ is positive definite \cite{kopp_supplemental_2023}. As a consequence, the feedback slows down the {\em instantaneous} velocity $\veci{v}(t)$, while there is an {\em aligning} effect with respect to the retarded velocity $\bm{v}_i(t-\tau)$. 
Deterministically, these effects persist  until the velocity vector becomes constant
(and, thus, $\bm{v}_i(t)=\bm{v}_i(t-\tau)$), yielding - for a single particle - a marginally stable long-time state \cite{kopp_persistent_2023-1}. Numerical results for the full, interacting system and a single realization of noise
within the velocity-aligned regime, are plotted in fig.~\ref{fig:mechanism}. 
Specifically, fig.~\ref{fig:mechanism}a) shows the (scalar) quantity $ N^{-1}\sum_i\bm{v}_i\cdot\bm{E}(\bm{R}_i)\left(\bm{v}_i(t) - \bm{v}_i(t-\tau)\right)$,
that represents the (system-averaged) projection of the feedback term appearing in eq.~(\ref{eq:vdot}) onto each $\bm{v}_i$. 
After a transient regime, the values are indeed positive everywhere, confirming the effects described before.
At longer times, the entire term approaches zero reflecting that the differences $\bm{v}_i(t) - \bm{v}_i(t-\tau)$ become small even in presence of interactions.

Consider now the second term on the r.h.s. of eq.~(\ref{eq:vdot}) involving the (configuration-dependent) matrix $\bm{A}(\bm{r}_{ij})$ that is defined via derivatives of the repulsive (WCA) force \cite{kopp_supplemental_2023}. This term
whose (short) range is determined by that of the potential $U^{\mathrm{WCA}}$,
couples the (instantaneous) velocity vectors of {\em different} neighboring particles $i$ and $j$. We note that the same type of (pair) coupling term appears within a corresponding analysis of interacting ABPs \cite{caprini_spontaneous_2020,caprini_spatial_2021-1, caprini_flocking_2023-1}, independent of the actual shape of the isotropic potential.
Due to the appearance of the difference $\bm{v}_i-\bm{v}_j$, the pair term naturally decouples into two parts \cite{kopp_supplemental_2023}. The first (``self'') part involving only $\bm{v}_i$ may be seen as an effective friction 
due to the presence of other particles. The second (``distinct'') part involves the actual coupling to the velocities $\bm{v}_j$ of other particles.
In particular, if the elements of $\bm{A}(\bm{r}_{ij})$ are positive, $\bm{v}_i$ tends to {\em align} with $\bm{v}_j$, providing a mechanism of velocity alignment reminiscent of the Vicsek model \cite{vicsek_novel_1995}.
Numerical results for the elements of the system-averaged matrix $\bm{A}=N^{-1}\sum_i\bm{A}_i$ with $\bm{A}_i=\sum_{j\neq i}\bm{A}(\bm{r}_{ij})$ are shown in fig.~\ref{fig:mechanism}b). We find that the matrix is essentially diagonal and that the diagonal elements are indeed positive. Also shown are corresponding predictions of a mean-field approximation \cite{kopp_supplemental_2023}, taking into account the fact that the structure in the ordered state is quite homogeneous (at least at large $\phi$).
In this case, $\bm{A}_i$ becomes particle-independent, that is, $\bm{A}_i=\bm{A}$. Assuming weak spatial correlations the approximation then yields a diagonal matrix with elements $\bm{A}_{xx(yy)}\approx \bar{A}=\rho\pi\varepsilon\tilde{b}$, where $\rho=N/L^2$ is the density. Thus, $\bar{A}$ is positive; it increases linearly in the density $\rho$ (or in $\phi=\rho\pi R^2$)
and in the repulsion parameter $\varepsilon$, while $\tilde{b}$ is a \emph{positive} quantity given by derivatives of $U^{\mathrm{WCA}}$ \cite{kopp_supplemental_2023}. From fig.~\ref{fig:mechanism}b) we see
that $\bar{A}$ provides indeed a reasonable estimate of the numerical values. The connection with the Vicsek model can be made more explicit by introducing (following \cite{caprini_spontaneous_2020}) the average velocity 
$\bm{v}^{*}= z^{-1}\sum_{j}'\bm{v}_j$, where the sum extends only over the $z$ neighboring particles. With this, eq.~(\ref{eq:vdot}) can be written as \cite{kopp_supplemental_2023} 
\begin{eqnarray}
	\gamma \vecidot{v} &=& -\bm{E}(\bm{R}_i)\left(\bm{v}_i(t) - \bm{v}_i(t-\tau)\right)\nonumber\\
	& & -\bar{A}\left(\bm{v}_i(t) -\bm{v}^{*}\right)+\sum_{j\neq i}\color{black}\bm{A}(\bm{r}_{ij})\left(\bm{v}_j(t) -\bm{v}^{*}\right).
	\label{eq:vdot2}
\end{eqnarray}
The second term (with coupling constant $\bar{A}>0$) directly reflects an aligning effect of $\bm{v}_i$ with the neighbor velocity $\bm{v}^{*}$.
The last term (involving deviations of the neighbor velocities relative to $\bm{v}^{*}$) is practically negligible, as the numerical calculations show.%

\section{Emergence of long-range ordering}
\label{sec:long-range}
While the mechanism described above explains the \emph{local} velocity alignment between neighboring particles (which also occurs in systems of repulsive ABPs \cite{caprini_spontaneous_2020}), there remains the question of how the system develops long-range velocity ordering. How is the information about a particle's velocity orientation relative to those of its neighbors transported through the system? 
We propose the following (mean-field-like) argument. 
Assume that a particle interacts (sterically) with another one such that the velocity vectors of the two particles align. This will influence the subsequent displacements of each particle that enter, by construction, the particle's histories and, thus, their time-delayed feedback forces (for an illustration, see SI \cite{kopp_supplemental_2023}). In this way,
the velocity directions are affected not only instantaneously (as in ABPs), but at least over the next delay time interval, $\tau$.
As a result, the particles can travel along the same direction even without having contact.
Moreover, if these particles collide with further ones within the delay time $\tau$, these further particles will be ''pushed'' and align their velocity directions along the same direction. In this way,
a loosely bounded ensemble of particles moving along the same direction can form.

These considerations imply that, for transport of ''alignment information'', the distance $l_\tau$ that a particle travels in one delay time,
needs to be comparable or larger than the typical distance between particles, $l_\mathrm{av}$.
For the most simple estimate of $l_\tau$, we use the (analytically accessible) value for a single, deterministic particle moving in one dimension, yielding $l_\tau=|\bm{r}(t)-\bm{r}(t-\tau)|=v_\infty\tau$ (in the long-time limit) \cite{kopp_persistent_2023-1}.
Please note that $l_\tau$ increases with $A/k_BT$ (recall that $v_\infty=(\sqrt{2}b/\tau)\sqrt{-\ln\left(\gamma b^2/(A\tau)\right)}$).
To estimate $l_\mathrm{av}$, we use the ideal-gas result $l_{\mathrm{av}}=\sqrt{1/\rho}$. At the onset of order, we 
expect the two lengths to be comparable. Setting $l_\tau=l_\mathrm{av}$ and solving with respect to the feedback strength we obtain 
\begin{equation}
	\frac{A^\mathrm{MF}}{k_BT}= \frac{\tau_B}{\tau}\exp\left[\frac{1}{8R^2\rho}\right]=\frac{A^\mathrm{min}}{k_B T}\exp\left[\frac{1}{8R^2\rho}\right],
	\label{eq:onset}
\end{equation}
where we recall that $A^\mathrm{min}/k_BT=\tau_B/\tau$ is the threshold value of persistent motion of a single particle. From the state diagram in fig.~\ref{fig:state_diag} it is seen that eq.~(\ref{eq:onset}) provides a reasonable estimate for the onset of ordering. The above prediction also shows directly that $A^\mathrm{MF}/k_BT\rightarrow A^\mathrm{min}/k_BT$ in the limit of large densities, consistent with our numerical data.

Clearly, the above argument is rather simplistic; it neglects, e.g., the impact of noise.
A more refined argument, closer in spirit to the Vicsek model \cite{vicsek_novel_1995,chate_collective_2008}, is based on the particle's \emph{persistence length} $l_p$,  i.e., the distance after which the direction of motion becomes uncorrelated due to thermal noise. To allow for an ordered state,
this length should be larger than the average length $l_\mathrm{mfp}$ (mean free path), after which two particles collide. Here we estimate $l_p$ by that of an \emph{isolated} feedback-driven particle, $l_p=v_{\mathrm{eff}}\tau_{\mathrm{eff}}$, where the two factors can be extracted from single-particle calculations as described in \cite{kopp_persistent_2023-1, kopp_supplemental_2023}. $l_p$ increases monotonically with the feedback strength (see SI \cite{kopp_supplemental_2023}). To estimate 
$l_\mathrm{mfp}$ as function of density, we perform simulations for the passive case, $A/k_BT=0$.
In fig.~\ref{fig:mechanism}c) we plot the order parameter $\langle |\bm{v}_{\mathrm{av}}|\rangle/v_\infty$
as function of $l_p/l_\mathrm{mfp}$ for different packing fractions. For all $\phi$ considered, the order parameter reaches appreciable values when the persistence length exceeds two or three times the length $l_\mathrm{mfp}$ (a corresponding argument holds for the related time scales, see SI). Similar behavior is seen if we further approximate $l_\mathrm{mfp}\sim\rho^{-1/2}$.
This relation between $l_p$ and $\rho$ at the onset of ordering is indeed reminiscent to what is seen in the Vicsek model \cite{chate_collective_2008}.
Moreover, we find from fig.~\ref{fig:mechanism}c) that the numerical data for different $\phi$ \emph{collapse}, indicating that the ratio  $l_p/l_\mathrm{mfp}$ indeed crucially determines the degree of ordering.
\section{Conclusions}
As our main result, we have reported the emergence of large-scale velocity alignment in a feedback-driven system
for a broad range of feedback parameters and densities. Within this state, the particles collectively move along one direction. A crucial ingredient for velocity alignment in our system is, first, 
the presence of steric interactions yielding clustering and local alignment. Different to the ABP system taken here as a reference, the long-time stability of the formed clusters is suppressed by history effects. 
A second major ingredient is the history dependence of the feedback forces. Thereby, steric interactions do not only lead to instantaneous alignment of velocities of neighboring particles, but rather affect 
the direction of the feedback forces and resulting propulsion over long times and distances, yielding eventually large-scale ordering.

Our predictions could be studied using an ensemble of colloidal particles which can be individually driven by optical forces \cite{qian_harnessing_2013,franzl_active_2020,loffler_behavior-dependent_2021}.
Another realization of our system might be an ensemble of biological ``particles'', particularly cells, on viscoelastic substrates where similar feedback effects due to the retarded response in the substrate occur \cite{clark_self-generated_2022}.

There are several open questions directly related to this study. This concerns, first, the nature of the transitions between the states observed here, including a more detailed study of the banding state seen especially at low densities and close above the 
threshold of ordering.
Indeed, based on studies on the Vicsek model \cite{chate_collective_2008}, one expects these questions to be accessible only in simulations of much larger systems than those considered here.
Second, it would be very interesting to explore differences between the present, artificially imposed delay effects and the impact of {\em inertial} delay. In fact, the latter has also shown to destabilize clustering and MIPS due to a ``bounce-back'' effect \cite{lowen_inertial_2020}.
Third, it seems promising to apply the present control scheme to active (rather than passive) Brownian particles. Indeed,
the study of active particles with retardation effects due
to feedback, inertia or other mechanisms is currently an emerging field \cite{mijalkov_engineering_2016, holubec_finite-size_2021-1, sprenger_active_2022, wang_spontaneous_2023,muinos-landin_reinforcement_2021}. From a more general perspective, our work contributes to establishing a link between collective systems under time-delayed feedback and active matter \cite{loos_irreversibility_2020}. Both types of systems are intrinsically out of equilibrium and, thereby, capable of self-organization. Exploring these connections may help us to better understand the autonomous dynamics of real systems of motile agents (such as bacteria, birds, and even humans) involving perception of information. In such systems, time delay is an essentially inevitable ingredient.
\acknowledgments
	We gratefully acknowledge the support of the Deutsche Forschungsgemeinschaft (DFG, German Research Foundation), project number 163436311 - SFB 910.
\bibliographystyle{eplbib}
\bibliography{MPPaperReferences}
%
\end{document}